\documentclass[preprint,showpacs,preprintnumbers,amsmath,amssymb]{revtex4}
\usepackage{graphicx}
\usepackage{dcolumn}
\usepackage{bm}

\begin{document}


\title{Interactions of Relativistic $^6$Li Nuclei with Photoemulsion Nuclei}

\author{M.~I.~Adamovich}
   \affiliation{Lebedev Institute of Physics, Russian Academy of Sciences, Leninskii pr. 53, Moscow, 117924 Russia} 
\author{I.~A.~Konorov}
   \affiliation{Lebedev Institute of Physics, Russian Academy of Sciences, Leninskii pr. 53, Moscow, 117924 Russia}
\author{V.~G.~Larionova}
   \affiliation{Lebedev Institute of Physics, Russian Academy of Sciences, Leninskii pr. 53, Moscow, 117924 Russia}
\author{N.~G.~Peresadko}
   \affiliation{Lebedev Institute of Physics, Russian Academy of Sciences, Leninskii pr. 53, Moscow, 117924 Russia}
\author{S.~P.~Kharlamov}
   \affiliation{Lebedev Institute of Physics, Russian Academy of Sciences, Leninskii pr. 53, Moscow, 117924 Russia}
\author{V.~G.~Bogdanov}
   \affiliation{Khlopin Radium Institute, St. Petersburg, 198021 Russia}
\author{V.~A.~Plyushchev}
   \affiliation{Khlopin Radium Institute, St. Petersburg, 198021 Russia}
\author{Z.~I.~Solovyeva}
   \affiliation{Khlopin Radium Institute, St. Petersburg, 198021 Russia}
   

\begin{abstract}
\indent 
Inelastic interactions of nuclei accelerated to a momentum of 4.5~GeV/$c$ per projectile nucleon with photoemulsion nuclei have been investigated. The main features of these interactions — mean ranges of $^6$Li nuclei, mean multiplicities of secondaries, the isotopic composition of fragments, fragmentation channels, and the mean transverse momenta of projectile fragments — have been measured. The probability of the charge-exchange reaction featuring lithium nuclei has been determined. The results obtained for the $^6$Li nucleus have been compared with data for other nuclei. The observed features of $^6$Li interactions with other nuclei indicate that the $^6$Li structure in the form of the loosely bound system consisting of an $\alpha$-particle and a deuteron cluster clearly manifests itself in these interactions. Events resulting in the coherent dissociation of $^6$Li nuclei into $^4$He+$d$, $^3$He+$t$, and $t+d+p$ and involving low-lying excitations of $^6$Li have been observed.\par
\indent \par
\indent DOI: 10.1134/S1063778810120161\par
\end{abstract}


\maketitle

\section{\label{sec:level1}INTRODUCTION}

\indent This article reports on an investigation of $^6$Li interactions with photoemulsion nuclei. The present experiment extends studies in the as-yet-unexplored interval of projectile masses by means of nuclear beams accelerated to a momentum of 4.5~GeV/$c$ per projectile nucleon at the synchrophasotron installed at JINR (Dubna). Here, we inquire into multiparticle-production processes. Specifically, we analyze in detail the fragmentation of $^6$Li projectiles into nuclei of helium and hydrogen isotopes. Investigation of the fragmentation of relativistic nuclei supplements classical experiments devoted to the disintegration of the same nuclei used as targets. In the present case, the detection threshold is close to zero, which makes it possible to study the process in question at very low energy-momentum transfers. The method of nuclear photoemulsions is very efficient in studying phenomena like projectile fragmentation owing to a high resolution of emulsions and to the possibility of observing interaction events in 4$\pi$-geometry. That all charged secondaries can be recorded and identified permits studying the isotopic composition of fragments and projectile-fragmentation channels. Events in which the sum of the fragment mass numbers is equal to the projectile mass number form a special class. In these events, we record all secondaries and can determine both the transverse-momentum transfer to the nucleus and the fragment transverse momenta in the reference frame comoving with the nucleus undergoing fragmentation. If, in this case, the projectile charge is conserved, the dissociation channels are recorded, but, if the total fragment charge changes in relation to the initial value of the nuclear charge, we can single out charge-exchange transitions experienced by the projectile nucleus. In this study, we estimate the invariant mass of dissociation products and compare it with the levels of excited states in the projectile nucleus.\par

\section{\label{sec:level2}EXPERIMENTAL PROCEDURE}

\indent A stack formed by layers of BR-2 NIIKhIMFOTOPROEKT photoemulsion with a relativistic sensitivity was exposed to a beam of $^6$Li nuclei accelerated to a momentum of 4.5~GeV/$c$ per projectile nucleon at the synchrophasotron installed at JINR (Dubna). The layers of thickness about 600 $\mu$m had the dimensions of 10$\times$20~cm$^2$. During irradiation, the beam was directed parallel to the emulsion plane along its long side. Events of interest were sought by means of scanning along the tracks; this allowed us to determine the mean range of $^6$Li nuclei with respect to inelastic interactions in photoemulsion. In the events being investigated, we measured the polar and azimuthal emission angles for all charged particles. The charges of relativistic particles were determined by the track-ionization density. In interactions of an accelerated $^6$Li nucleus with photoemulsion nuclei, the charge of a relativistic secondary does not exceed three, the determination of such charges being highly reliable in the emulsion that we used. The momenta of singly and doubly charged relativistic particles with emission angles not exceeding 3$^\circ$ were estimated on the basis of measurements of multiple Coulomb scattering of the tracks. Relying on the results of these measurements, we identified the hydrogen and helium isotopes in the composition of $^6$Li fragments.\par
 	
\section{\label{sec:level3}MEAN RANGE OF $^6$Li WITH RESPECT TO INTERACTIONS}
	
\begin{table}
\caption{\label{Table:1} Mean range $\lambda$ with respect to the inelastic interactions of nuclei in photoemulsion}
\label{Table:1}       
\begin{tabular}{c|c|c|c|c}
\hline\noalign{\smallskip}
Projectile & Momentum per projectile & \multicolumn{2}{c|}{$\lambda$, cm} &\\
nucleus & nucleon, GeV/$c$ & calculated & experimental & References \\
\noalign{\smallskip}\hline\noalign{\smallskip}
$^4$He & 4.5 & 19.6 & $19.5\pm0.3$ & \cite{Tolstov} \\
$^6$Li & 4.5 & 16.5 & $14.1\pm0.4$ & Our study \\
$^{12}$C & 4.5 & 13.5 & $13.7\pm0.5$ & \cite{Bannik} \\
$^{16}$O & 4.5 & 12.1 & $13.0\pm0.5$ & \cite{Bannik2} \\
$^{22}$Ne & 4.1 & 10.6 & $10.2\pm0.1$ & \cite{Bannik2} \\
$^{24}$Mg & 4.5 & 10.0 & $9.6\pm0.4$ & \cite{Sharkawy} \\
\noalign{\smallskip}\hline
\end{tabular}
\end{table}

\indent Over the total length of scanned tracks, which was equal to 234.1~m, we recorded 1657 inelastic interactions of $^6$Li with the nuclei of elements entering into the composition of the emulsion used. Thus, the mean range of $^6$Li with respect to interactions in photoemulsion was $\lambda_{Li}=14.13\pm0.35$~cm. This value and the values of $\lambda_{A}$ in photoemulsion that were obtained in \cite{Tolstov,Bannik,Bannik2,Sharkawy} for some other projectile nuclei with a momentum of 4.5~GeV/$c$ per nucleon are presented in Table~1. Also quoted in this table are the values calculated by the formula $\lambda_{A_p}=1/\sum_{t}N_t\sigma_{A_pA_t}$, where $A_p$ and $A_t$ are, respectively, the projectile and target mass numbers; $\sigma_{A_pA_t}$ is the cross section for projectile-target interaction; and $N_t$ is the concentration of $A_t$ nuclei in the emulsion. The cross sections were calculated within geometric overlap model by the Bradt-Peters formula \cite{Bradt} 
$$
\sigma_{A_pA_t}~=~\pi r^2(A_p^{1/3}~+~A_t^{1/3}~-~b)^2,
$$
where $r=1.23$~fm and the overlap parameter is $b=1.56-0.2(A_p^{1/3}+A_t^{1/3})$. This approximation satisfactorily describes experimental data in a broad interval of projectile mass numbers. It can be seen that, for the nuclei presented in Table 1, the calculated values comply fairly well with experimental data. The $\lambda_{Li}$ value that we obtained is substantially smaller than the calculated one. In Fig.~1, it can be seen more clearly that, according to this model, the nucleus with mass number $A=11\pm1$ corresponds to the experimental value of $\lambda_{Li}$. This indicates that the $^6$Li nucleus has an anomalously large radius, in agreement with the conclusions drawn in \cite{Bayman}. At the same time, the corresponding nucleon-distribution radius, 2.7$\pm$0.1~fm, which determines the cross sections for inelastic nucleus-nucleus interactions, exceeds the charge-distribution radius in the $^6$Li nucleus.\par

\begin{figure}
    \includegraphics[width=4in]{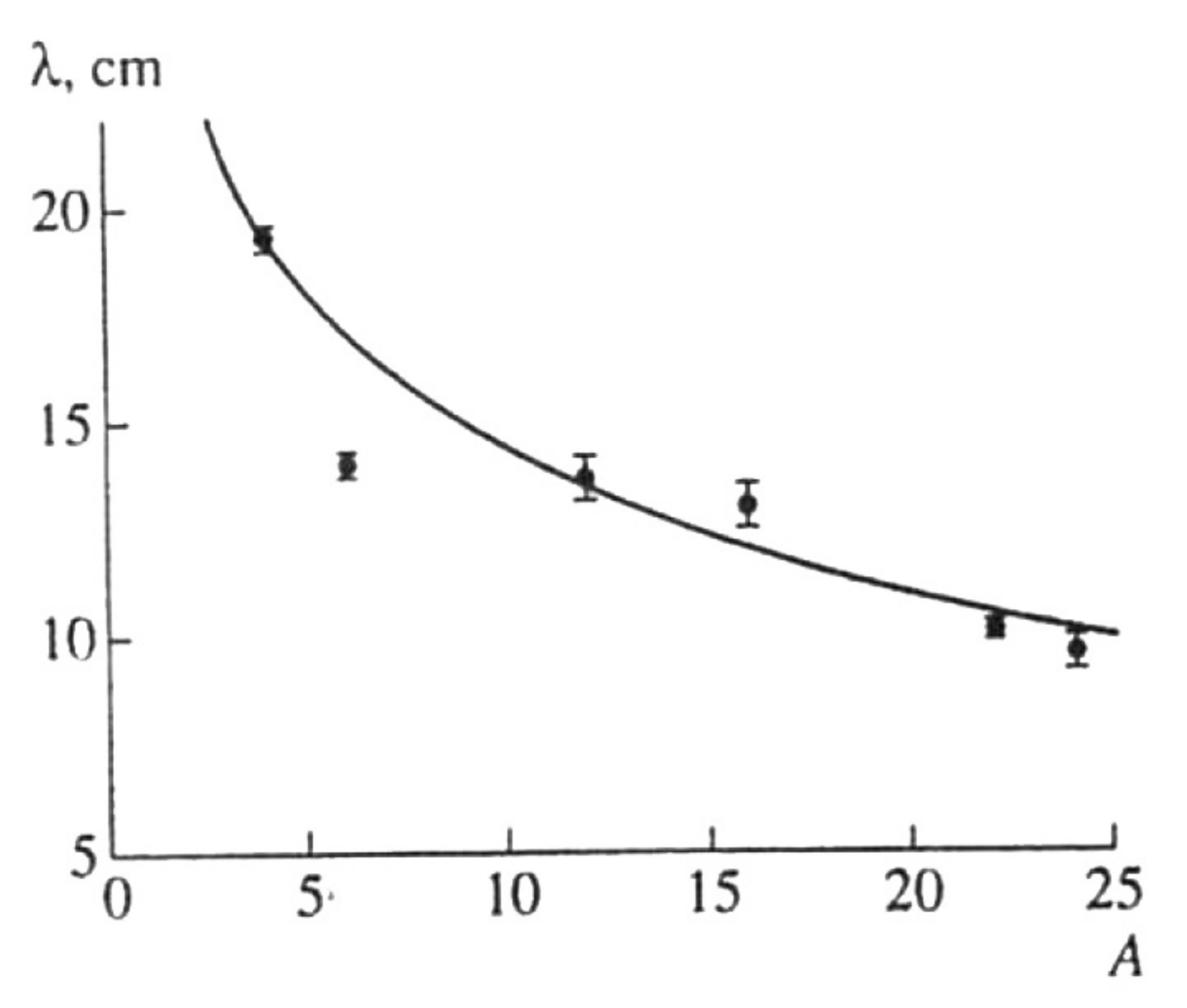}
    \caption{\label{Fig:1} Mean range $\lambda$ of a projectile with respect to inelastic interactions in the photoemulsion as a function of the projectile mass number. The curve represents a fit obtained within the geometric model.}
    \end{figure}

\section{\label{sec:level4}MULTIPLICITIES OF SECONDARY PARTICLES}
	
\indent In 1014 events, we measured the emission angles for all charged particles. Taking into account the resulting characteristics, we divided all charged particles according to the scheme adopted in the emulsion procedure.\par

\indent As fragments of a projectile $^6$Li nucleus, we classify particles that conserved the primary-nucleus velocity of $\beta\sim0.98$~that is, those that did not suffer inelastic interactions with the target. For the most part, they are contained in a narrow forward cone of the polar angle $\theta$, which is estimated as $\theta=0.2/p_0$, where the factor 0.2~GeV/$c$ is determined by the interval of spectator-nucleon transverse momenta, while $p_0$ is the momentum of the accelerated-projectile nucleon. At $p_0$=4.5~GeV/$c$, the opening angle of the fragmentation cone is about 3$^{\circ}$. In order to identify the fragments and to determine their masses, we measured the momenta of the particles whose emission angles were within this cone. We distinguished\par

\indent (i)	$s$ particles, singly charged relativistic particles with velocities $\beta>0.7$ and relative ionization $I/I_0<1.4$, where $I_0$ is the particle track density at the minimum of the ionization curve (predominantly, these are product mesons and inelastically interacted protons with the emission angles larger than the fragmentation-cone angle);\par

\indent (ii) $g$ particles, fast target fragments with relative ionization $I/Io>1.4$ and mean range in excess of 3~mm (mainly, these are protons with energies $E_p~>~26$~MeV);\par

\indent (iii) $b$ particles, slow target fragments with mean range less than 3~mm; and\par

\indent (iv) $h$ particles, the group of all target fragments including $g$ and $b$ particles.\par

\indent In Fig. 2, the mean multiplicities of particles for $^6$Li nuclei are contrasted against experimental data from \cite{Bannik,Bannik2,Sharkawy,Heckman,Bogdanov,Bubnov}. Within errors, the values of $\langle n_s\rangle$ obtained for $^6$Li nuclei are in accord with the curve describing $\langle n_s\rangle$ as a function of the projectile mass. The value of $\langle N_h\rangle$ for $^6$Li nuclei does not fit in the general trend observed in the interactions of nuclei investigated previously, falling well below the curve that describes the aforementioned dependence. This suggests a larger contribution of peripheral collisions in $^6$Li interactions than in the interactions of other nuclei.\par

\begin{figure}
    \includegraphics[width=4in]{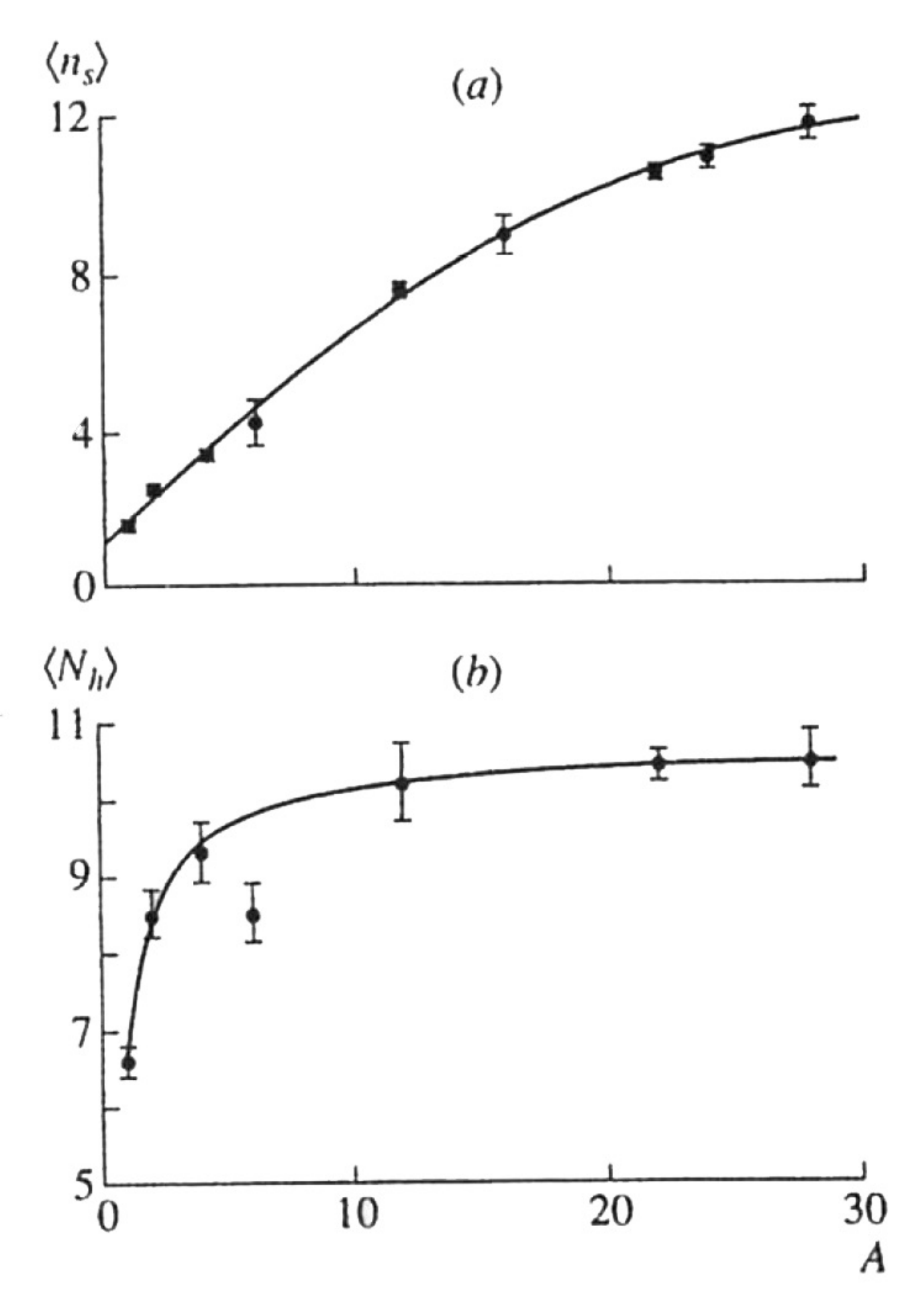}
    \caption{\label{Fig:2} Mean multiplicities of secondaries versus the projectile mass number: ($a$) results for singly charged relativistic particles, $\langle n_s\rangle$, and ($b$) results for target fragments, $\langle N_h\rangle$. The curves represent smoothed polynomial fits.}
    \end{figure}

\indent Let us now consider the multiplicities of particles of the various kinds versus the character of projectile fragmentation. In order to characterize nuclear fragmentation, we use the total charge of the fragments, $Q=\sum Z_f$, where $Z_f$ stands for the charges of individual projectile-fragment species. The value of $Q$ can be used to estimate the mean number $v$ of inelastically interacted nucleons of the projectile nucleus. This can be done by using the expression $v=2(Z - Q)$. Events for which $Q=0$ are usually referred to as those of central (or head-on) collisions. The set of $Q=3$ events includes inelastic interactions preserving the projectile nucleus, events featuring the diffractive and electromagnetic dissociation of the projectile nucleus, and events that are due to the interaction of projectile neutrons with target nuclei. Table 2 presents the mean multiplicities of secondaries versus $Q$ and $v$ for accelerated $^6$Li nuclei, deuterons, and alpha particles. Also quoted in this table are the mean multiplicities of secondaries originating from proton-nucleus interactions in photoemulsion at a momentum of 4.5~GeV/$c$ per projectile nucleon. For all nuclei, the particle multiplicities display a tendency to grow with increasing $v$. It should be emphasized that, at equal nonzero values of $v$, the values $\langle n_s\rangle$ are approximately equal for all the aforementioned nuclear species. Moreover, the specific multiplicities of $s$ particles, $\langle n_s\rangle/v$, are consistent within errors. Thus, multiparticle-production processes proceed in the same way in all these events, 1.4 $s$ particles being observed on average in an inelastic interaction of a projectile nucleon. The total number of particles per event is determined by the number of inelastically interacted projectile nucleons. In the set of $v=0$ events, the values $\langle n_s\rangle$ can be used to estimate the number of inelastically interacted projectile neutrons. At the same time, the mean multiplicities of nonrelativistic particles at fixed è depend on the projectile mass. With increasing projectile mass, the multiplicity of target fragments decreases; that is, the relative contribution of peripheral interactions increases.\par

\begin{table}
\caption{\label{Table:2} Mean multiplicities of secondaries for various values of the parameters $Q$ and $v$}
\label{Table:2}       
\begin{tabular}{c|c|c|c|c|c|c}
\hline\noalign{\smallskip}
Projectile nucleus&$Q$&~~$v$~~&$\langle n_s\rangle$&$\langle n_b\rangle$&$\langle n_g\rangle$&$\langle N_h\rangle$\\
\noalign{\smallskip}\hline\noalign{\smallskip}
$^6$Li & 0 & 6 & $10.2\pm1.0$ & $11.8\pm1.7$&$8.3\pm0.4$&$20.1\pm2.0$ \\
$^6$Li & 1 & 4 & $6.2\pm0.5$ & $5.8\pm0.9$&$5.5\pm0.2$&$11.3\pm1.2$ \\
$^4$He & 0 & 4 & $5.6\pm0.1$ & $7.2\pm0.3$&$6.8\pm0.2$&$14.0\pm0.5$ \\
$^6$Li & 2 & 2 & $2.7\pm0.3$ & $2.7\pm0.4$&$2.9\pm0.2$&$5.5\pm0.3$ \\
$^4$He & 1 & 2 & $2.6\pm0.1$ & $2.4\pm0.2$&$2.7\pm0.2$&$5.1\pm0.3$ \\
$^2$H  & 0 & 2 & $2.4\pm0.1$ & $4.4\pm0.3$&$4.9\pm0.2$&$9.3\pm0.5$ \\
$^6$Li & 3 & 0 & $0.8\pm0.1$ & $0.9\pm0.2$&$1.2\pm0.1$&$2.0\pm0.2$ \\
$^4$He & 2 & 0 & $1.4\pm0.1$ & $1.6\pm0.2$&$2.0\pm0.2$&$3.6\pm0.4$ \\
$^2$H  & 1 & 0 & $0.7\pm0.1$ & $3.0\pm0.3$&$3.5\pm0.3$&$6.4\pm0.5$ \\
$^1$H  &   &   & $1.6\pm0.1$ & $2.8\pm0.1$&$3.8\pm0.1$&$6.6\pm0.1$ \\
$^6$Li & All $Q$ &  & $4.3\pm0.6$ & $4.6\pm0.6$&$4.0\pm0.1$&$8.5\pm0.6$ \\
\noalign{\smallskip}\hline
\end{tabular}
\end{table}

\indent By convention, the nuclear composition of photoemulsion is divided into three groups. The first group contains nuclei of free hydrogen. Carbon, nitrogen, and oxygen form the group of light nuclei, while silver and bromine constitute the group of heavy nuclei. The number of particles appearing to be target fragments, $N_h=n_b+n_g$, indicates the group of nuclei on which a given interaction event occurred. In accordance with this, it is customary to consider the following groups of events:\par
	
\indent (i) quasinucleon events with $N_h=0$~and 1; \par

\indent (ii) interactions with light nuclei (C, N, and O) and peripheral interactions with Ag and Br nuclei ($N_h=2-6$); and\par

\indent (iii) events with $N_h>7$, which form the class of nonperipheral interactions with Ag and Br nuclei.\par

\indent Apart from this, we single out $N_h>28$~events, which result in the complete breakup of Ag and Br nuclei. These events constitute $(6.7\pm2.0)\%$ of all measured interactions. Since the fraction of interactions featuring the heavy nuclei of the photoemulsion is about 66\% \cite{Galstyan}, the probability of the complete breakup of Ag and Br nuclei is $(10\pm3)\%$. Within uncertainties, this result agrees with the calculated value of 8.5\% obtained from the analysis of the breakup cross sections in \cite{Bannik3}.\par

\indent Table 3 presents the mean multiplicities of $s$, $g$, and $b$ particles for each group of events. The growth of the multiplicities of all kinds of particles with increasing $N_h$ is due to an increase in the amount of target nuclear matter on the way of the projectile nucleus. Also displayed in this table are the mean number of the inelastically interacted projectile nucleons, $\langle v\rangle$, and the specific multiplicity $\langle n_s/v\rangle$, which can be considered as an estimate of the multiplicity of inelastic interactions between projectile and target nucleons. It can be seen that an increase in $N_h$ leads to the growth of both the number of interacted projectile nucleons and the number of interactions suffered by each projectile nucleon on its target counterparts. In 60\% of $N_h\le1$ events, the total charge is conserved in the fragmentation cone. More than 70\% of all $N_h>28$~events do not involve charged projectile fragments; that is, a maximal number of projectile nucleons undergo inelastic collisions in these events.\par

\begin{table}
\caption{\label{Table:3} Moan multiplicities of secondaries in $^6$Li interactions with photoemulsion nuclei in various $N_h$ groups}
\label{Table:3}       
\begin{tabular}{c|c|c|c|c|c|c}
\hline\noalign{\smallskip}
$N_h$&$\langle n_s\rangle$&$\langle v\rangle$&$\langle n_s/v\rangle$&$\langle n_g\rangle$&$\langle n_b\rangle$&Number of events\\
\noalign{\smallskip}\hline\noalign{\smallskip}
0-1 & $1.3\pm0.3$ & $1.0\pm0.1$ & $0.9\pm0.1$ & $0.3\pm0.1$ & $0.14\pm0.02$ & 294 \\
2-6 & $3.5\pm0.7$ & $2.4\pm0.2$ & $1.4\pm0.3$ & $1.9\pm0.2$ & $2.1\pm0.2$ & 309 \\
$\ge7$ & $.2\pm1.2$ & $4.1\pm0.2$ & $1.7\pm0.2$ & $9.6\pm0.5$ & $8.2\pm0.6$ & 411 \\
$\ge28$ & $11.1\pm0.7$ & $5.4\pm0.1$ & $2.1\pm0.1$ & $20.3\pm1.0$ & $12.8\pm1.0$ & 68 \\
$\ge0$ & $4.3\pm0.6$ & $2.7\pm0.2$ & $1.4\pm0.1$ & $4.6\pm0.6$ & $4.0\pm0.1$ & 1014 \\
\noalign{\smallskip}\hline
\end{tabular}
\end{table}

\section{\label{sec:level5}ISOTOPIC COMPOSITION OF FRAGMENTS AND CHANNELS OF $^6$Li FRAGMENTATION}

\indent In order to determine the mass of a fragment of an accelerated nucleus, it is sufficient to know its momentum. In photoemulsions, the momentum of a particle is determined by measuring the multiple Coulomb scattering of this particle. The mean deviation $|D|$ of the track over a cell of length $t$ is related to the quantity $p\beta c$ by the equation $\langle|D|\rangle=(Z_fKt^{3/2})/(573p\beta c)$, where $Z_f$, $p$, and $\beta c$ are the fragment charge, momentum, and velocity, respectively, while $Ê$ is the scattering constant, whose numerical value is well known. Distortions and spurious scattering were taken into account within the so-called $p$ method \cite{Voinov}. In the case of multiple Coulomb scattering, the distribution of $\langle|D|\rangle$ for individual particles with identical charges and momenta must be close to a normal distribution. For fragments with identical velocities and identical charges, but with different masses, the 1/($ð\beta c)$ distribution must represent a superposition of a few normal distributions. At the same time, the more informative distributions of particles with respect to $p\beta c$ are used to perform a mass separation of nuclei. We will consider both these methods for the mass separation of particles and compare the results that these methods produce.\par

\begin{figure}
    \includegraphics[width=4in]{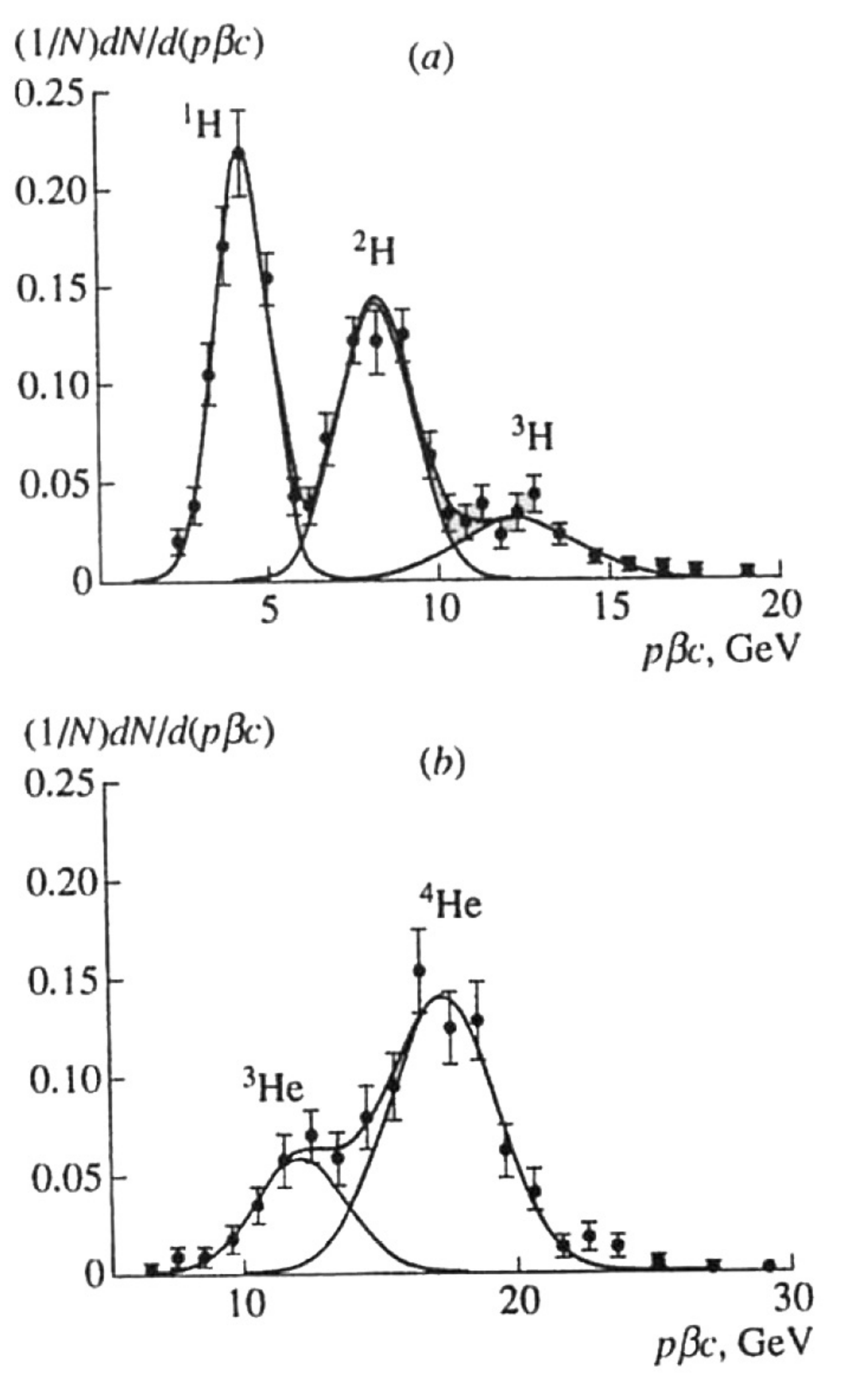}
    \caption{\label{Fig:3} Density $(1/N)dN/d(p\beta c)$ as a function of $p\beta c$: ($a$) results for singly charged $^6$Li fragments and ($b$) results for doubly charged $^6$Li fragments. The curves represent least squares fits in terms of Gaussian distributions (see main body of the text).}
    \end{figure}

\indent In Fig. 3, the results of measurements for the multiple scattering of particles are displayed as $p\beta c$ distributions of the normalized density $(1/N)dN/d(p\beta c)$. The mean error in the result¬ing values of $p\beta c$ is about 15\%. Figure 3$a$ shows the distribution of 892 singly charged relativistic particles recorded experimentally. In the region $2<p\beta c<17$~GeV, the least squares fit of experimental values to the sum of three normal distributions is quite satisfactory ($\chi^2=0.9$~per degree of freedom). The curves in Fig. 3 represent both the total distribution and its individual terms. The maxima of the approximating distributions are located at pfic values equal to 4.2, 8.2, and 12.2~GeV. These are sufficiently close to integral multiples of $p_0\beta c$, where $p_0$ is the projectile-nucleon momentum, and correspond to the $^1$H, $^2$H, and $^3$H isotopes. The variances of the proton and deuteron distributions are about 1~GeV, while the variance of the triton distribution is about 2~GeV. In passing, we note that the neighboring distributions overlap. The boundaries of fragment separation in $p\beta c$ were chosen in such a way that the overlapping areas of each of the distributions beyond the boundary were equal. The protons (with $p\beta c<6.0$~GeV), whose overlap with the deuterons was less than 2\%, were separated best of all. A markedly greater uncertainty in estimating the fragment mass is observed near the boundary separating deuterons and tritons ($p\beta c=10.0$~GeV), where about 4\% of the deuteron distribution and 12\% of the triton distribution overlap. Figure 3$b$ shows the analogous distribution for the measured 340 doubly charged relativistic particles. In order to determine the boundary separating $^3$He and $^4$He fragments, the experimental data in the interval $8<p\beta c<22$~GeV were fitted to the sum of two normal distributions (with $\chi^2=0.7$ per degree of freedom). The maxima of the distributions occur at $p\beta c$ values of 12.1 and 17.2~GeV, which agree well with the values expected for $^3$He and $^4$He. The variances of the distributions are about 2~GeV. The boundary separating these fragments is situated at $p\beta c$=14.0~GeV, and overlapping areas of the distributions from the boundary are 13 and 5\% for $^3$He and $^4$He, respectively. A similar procedure of isotope separation among fragments was employed by Lepekhin $et~al.$ \cite{Lepekhin} in studying $^6$Li fragmentation. \par
	
\begin{figure}
    \includegraphics[width=4in]{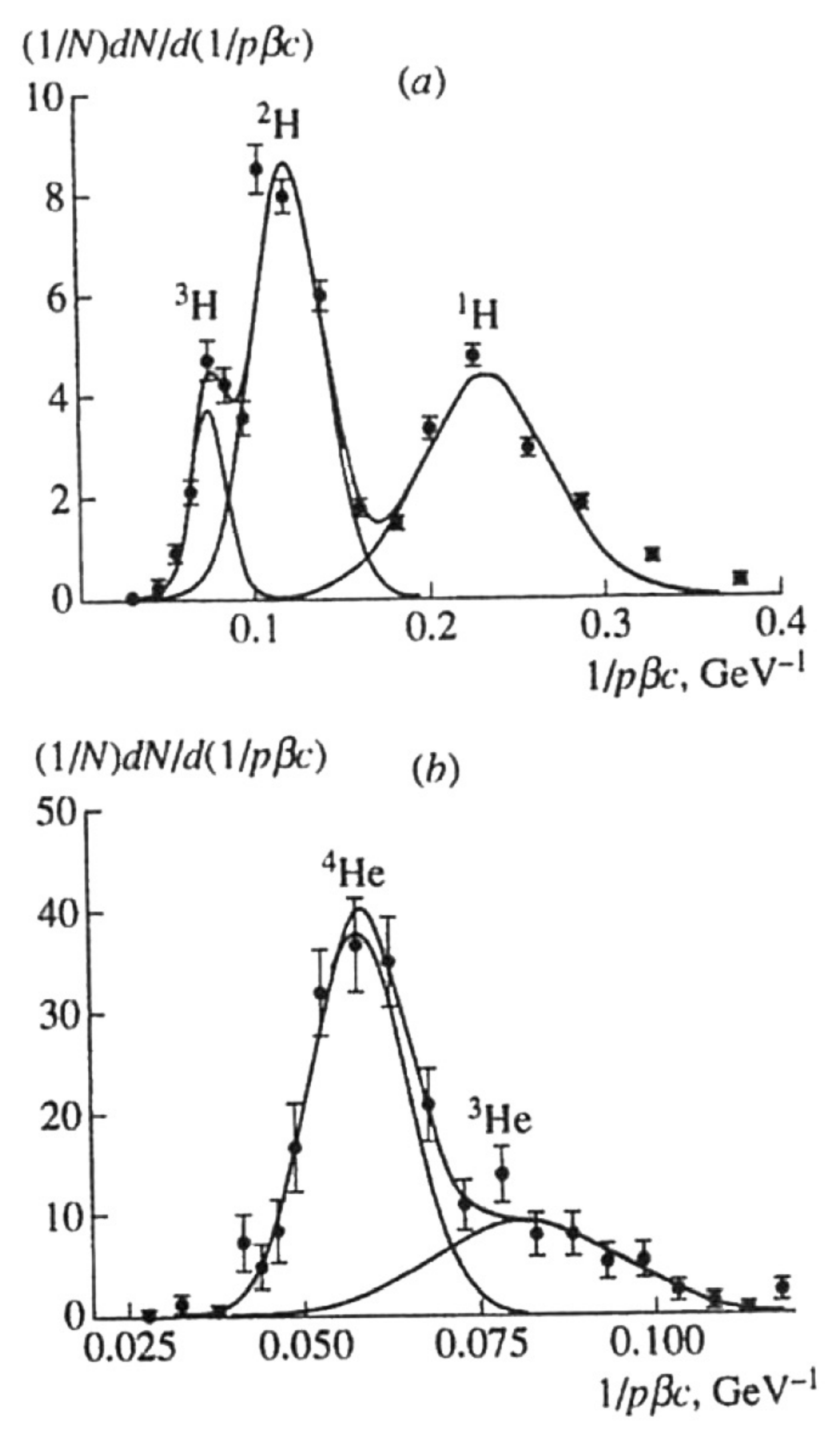}
    \caption{\label{Fig:4} Density $(1/N)dN/d(1/p\beta c)$ as a function of $1/p\beta c$: $(a)$ results for singly charged $^6$Li fragments and ($b$) results for doubly charged $^6$Li fragments. The curves represent least squares fits of the distributions (see main body of the text).}
    \end{figure}

\indent Let us now perform isotope separation by using the $1/p\beta c$ distributions of particles. Figure 4 displays normalized experimental distributions. In fitting our experimental data on singly charged fragments in the region $1/p\beta c<0.3$~GeV$^{-1}$ to the sum of three normal distributions, the resulting $\chi^2$ value was 1.5 per degree of freedom. The boundaries of isotope separation with respect to $1/p\beta c$ were determined in just the same way as in the preceding case. The boundary separating the protons and the deuterons corresponds to the value of $p\beta c=5.8$~GeV, which is virtually coincident with the value obtained from the $p\beta c$ distribution. The overlap of the triton and deuteron distributions in $1/p\beta c$ is much larger than the overlap of the corresponding distributions in $p\beta c$, and the resulting boundary between the tritons and the deuterons shifts to the value of $p\beta c=12.1$~GeV. In fitting the experimental $1/p\beta c$ distributions of doubly charged fragments to the sum of two normal distributions, the resulting value of $\chi^2$ was 0.8 per degree of freedom. The overlap of the $1/p\beta c$ distributions was somewhat larger than the overlap of the $p\beta c$ distributions, and the resulting boundary between $^3$He and $^4$He corresponded to the value of $p\beta c=15.2$~GeV.\par

\indent Comparing the above two descriptions of experimental data on the rescattering of projectile fragments, we can see that, in terms of the $\chi^2$ criterion, the description on the basis of $1/p\beta c$ does not have advantages over the description of data on the basis of $p\beta c$. At the same time, this comparison makes it possible to assess the accuracy to which we perform the mass separation of individual fragment species. In both cases, we have obtained complying conditions for proton separation and comparatively close results for $^3$He and $^4$He separation, but noticeable distinctions were observed in determining the boundaries separating tritons. The ultimate boundaries between the fragments were chosen by taking into account the $\chi^2$ values for the fitted distributions. We used the $p\beta c$ values of 5.9 and 10.7~GeV for singly charged isotopes and the $p\beta c$ value of 14.4~GeV for the boundary separating $^3$He and $^4$He. Doubly charged particles with $p\beta c>22$~GeV were identified as $^6$He only when events in which they were observed could be associated with $^6$Li charge exchange on $^6$He. The resulting data on the isotopic composition of fragments and data on the mean isotope multiplicities are presented in Table 4. The errors quoted in this table include not only statistical uncertainties but also errors in determining the boundaries separating isotopes. An extremely high yield of deuterons, which is virtually identical to the proton yield, is a feature peculiar to the isotopic composition of $^6$Li fragments, because no such feature has been observed in the fragmentation of $^4$He, $^{12}$C, $^{22}$Ne, and $^{28}$Si nuclei \cite{A,D,B}. The yields of $^3$H and $^3$He nuclei, which have identical mass numbers, were virtually coincident.\par

\begin{table}
\caption{\label{Table:4} Isotopic composition of $^6$Li fragments}
\label{Table:4}       
\begin{tabular}{c|c|c|c}
\hline\noalign{\smallskip}
Charge&Isotop&Composition, \%&Mean multiplicity\\
\noalign{\smallskip}\hline\noalign{\smallskip}
1 & $^1$H & $44\pm7$ & $0.40\pm0.07$ \\
  & $^2$H & $43\pm7$ & $0.39\pm0.05$ \\
  & $^3$H & $13\pm3$ & $0.11\pm0.03$ \\
  &  All  &          & $0.90\pm0.02$ \\
2 & $^3$He & $30\pm7$ & $0.11\pm0.02$ \\
  & $^4$He & $68\pm8$ & $0.25\pm0.03$ \\
  & $^6$He & $1.4\pm0.3$ & $0.005\pm0.002$ \\
  &  Alll  &   & $0.36\pm0.02$ \\
3 & $^6$Li &   & $0.13\pm0.06$ \\
\noalign{\smallskip}\hline
\end{tabular}
\end{table}

\indent In each individual event, the set of all fragments characterizes a channel of projectile fragmentation and is represented here as a topological formula for this event. We have found charged projectile fragments in 835 events. In 768 of these, we were able to determine the nature of all fragments involved in an individual event. The results are displayed in Table 5. It can be seen that more than 80\% of events feature at least one charged fragment and that all possible fragmentation channels are realized. The table also contains the number of inelastic interactions preserving the $^6$Li nucleus involved and the number of $Q=0$ events without charged fragments. Data on associated multiplicities of fragments make it possible to perform a detailed comparison of the production of fragments having identical mass numbers ($^3$H and $^3$He). Within statistical uncertainties, we observed identical yields of these fragments both in the case of single production and in the case of associated production (together with other fragments). Events where the sum of fragment mass numbers is equal to the projectile mass number are worthy of special note. In 61 events (about 6\% of all interactions), $^6$Li disintegrates into two fragments, $^6$Li$\rightarrow^4$He+$d$, the majority of these events being those of coherent dissociation; that is, these are events not accompanied by the production of relativistic particles. A large contribution of this process to the cross section indicates that the system consisting of an alpha-particle and a deuteron cluster has a large weight in the structure of the $^6$Li nucleus. We recorded eight events (this is about 0.8\% of the total number of events) of coherent dissociation of $^6$Li into $^3$He and $^3$H. The emergence of these fragments can be interpreted as a manifestation of another possible nuclear structure in the form of $^3$He and $^3$H clusters. We also detected eight events of dissociation into only singly charged fragments, $^6$Li$\rightarrow^3$H+$^2$H+$^1$H. The last channel can receive contributions from both the aforementioned structures. Thus, we can see that, in the dissociation of $^6$Li, the manifestation of the structure in the form of $^3$He and triton clusters is an order of magnitude weaker than the manifestation of the structure formed by an alpha-particle and a deuteron cluster. In view of this, the unusually large yield of deuterons can be explained by a substantial contribution of events in which individual loosely bound clusters of $^6$Li effectively interact with target nuclei.\par

\begin{table}
\caption{\label{Table:5} Channels of $^6$Li fragmentation}
\label{Table:5}       
\begin{tabular}{c|c|c|c|c}
\hline\noalign{\smallskip}
Charge& & &Charge&\\
particle&Number of& &particle&Number of\\
content of&events& &content of&events\\
a channel& & &a channel&\\
\noalign{\smallskip}\hline\noalign{\smallskip}
Q=0 & 179 &  & $tdp$ & 8 \\
 $p$ & 90 &  & $tt$ & 4 \\
$pp$ & 23 &  & $^3$He & 46 \\
$ppp$ & 5 &  & $^3$He$p$ & 18 \\
$pppp$ & 1 &  & $^3$He$pp$ & 1 \\
$d$ & 97 &  & $^3$He$d$ & 29 \\
$dp$ & 64 &  & $^3$He$dp$ & 1 \\
$dpp$ & 8 &  & $^3$He$t$ & 8 \\
$dd$ & 27 &  & $^4$He & 97 \\
$ddp$ & 8 &  & $^4$He$p$ & 65 \\
$ddd$ & 3 &  & $^4$He$pp$ & 4 \\
$t$ & 32 &  & $^4$He$d$ & 61 \\
$tp$ & 24 &  & $^6$He & 5 \\
$tpp$ & 5 &  & $^6$Li & 13 \\
$td$ & 21 &  &  &   \\
\noalign{\smallskip}\hline
\end{tabular}
\end{table}

\indent The probability that a $^6$Li nucleus is preserved (there are 13 events featuring a triply charged fragment) in an inelastic interaction with photoemulsion nuclei is moderately low [(1.3$\pm$0.6)\%], much less than the probability that a $^4$He projectile is preserved {according to \cite{B}, the latter is (3.5$\pm$0.5)\%}. This can also be explained by the loosely bound cluster structure of the $^6$Li nucleus.\par

\section{\label{sec:level6}TRANSVERSE MOMENTA OF $^6$Li FRAGMENTS}

\indent The transverse momenta of fragments in the laboratory frame were calculated according to the expression $p^A_t=p_0Asin\theta$, where $p_0$ is the momentum of a projectile nucleon, $A$ is the fragment mass number, and $\theta$ is the measured polar angle of fragment emission. The mean transverse momenta $\langle p^A_t\rangle$ of $^6$Li fragments are listed in Table 6. Also quoted in this table are the experimental values of the mean transverse momenta of $^4$He and $^{12}$C fragments for the same momentum per projectile nucleon. For singly charged $^6$Li fragments, $\langle p^A_t\rangle$ is seen to grow with increasing fragment mass. These values are intermediate between those for $^4$He and $^{12}$C fragments. The transverse momenta of the doubly charged $^6$Li fragments are considerably lower than those for neighboring nuclei. The degree to which our results can comply with the model description of the fragmentation process \cite{Goldhaber} can be assessed by comparing the estimated model variance $\sigma_0$ deduced from the experimental values of $\langle p^A_t\rangle$. In doing this, we use the model relation between $\sigma_0$ and the variances $\sigma^A$ of the transverse-momentum distributions of fragments and the relation between $\sigma^A$ and $\langle p^A_t\rangle$ determined by a Rayleigh distribution. In the case of this estimate, the experimental values of $\langle p^A_t\rangle$ for $^2$H, $^3$H, and $^4$He can be considered to be compatible with the value of $\sigma_0=100\pm10$~MeV/$c$. By using data for protons, we obtain a noticeably smaller value of 80$\pm$5~MeV/$c$; a still smaller value of 70$\pm$5~MeV/$c$ was deduced from data for $^3$He nuclei. In order to eliminate the possible effect of a $^4$He admixture on the last value, we considered the region $p\beta c<13$~GeV, where there is virtually no overlap with the $^4$He distribution. In this case, the value $\langle p_t\rangle$ does not change within errors. This suggests that the smaller value of $\langle p_t\rangle$ for $^3$He is not due to errors in separating $^3$He and $^4$He nuclei. The observed distinctions in the estimated values of $\sigma_0$ can be caused by the fact that the transverse momenta of all $^6$Li fragments cannot be described by a single distribution. A detailed analysis of the transverse momenta of projectile fragments can prove to be an efficient tool for studying inelastic nuclear interactions.\par

\begin{table}
\caption{\label{Table:6} Mean transverse momenta of nuclear fragments in the laboratory frame (in MeV/$c$)}
\label{Table:6}       
\begin{tabular}{c|c|c|c|c|c}
\hline\noalign{\smallskip}
Projectile nucleus& $^1$H & $^2$H & $^3$H & $^3$He & $^4$He\\
\noalign{\smallskip}\hline\noalign{\smallskip}
$^4$He & $86\pm3$ & $142\pm7$ & $175\pm14$ & $223\pm12$ & $239\pm20$ \\
$^6$Li & $97\pm10$ & $153\pm5$ & $189\pm17$ & $131\pm8$ & $139\pm8$ \\
$^{12}$C & $112\pm2$ & $203\pm10$ & $223\pm25$ & \multicolumn{2}{c}{$238\pm8^*$} \\
\noalign{\smallskip}\hline
\multicolumn{6}{l}{$^*$~In the present experiment, all doubly charged fragments were}\\
\multicolumn{6}{l}{assumed to be $^4$He nuclei.}
\end{tabular}
\end{table}

\section{\label{sec:level7}TRANSVERSE MOMENTA OF $^6$Li FRAGMENTS}

\indent In order to estimate the contribution of $^6$Li charge exchange resulting in $^6$He formation, we consider events in which doubly charged particles are characterized by $p\beta c$ values in excess of 22~GeV. In Fig. 36, we can see that the experimental pfic distribution in this region, which corresponds to $^6$He fragments, lies slightly above the approximating curve. In 12 events featuring such a doubly charged fragment, there are no other projectile fragments. Of these, only five events containing one singly charged relativistic particle each can be associated with the inelastic process $^6$Li$\rightarrow^6$He+$\pi^+$. The remaining seven events involve a few singly charged relativistic particles each. Assuming that charge-exchange processes and processes featuring the production of a few mesons have considerably smaller cross sections, we associated the doubly charged fragments in these events with $^4$He nuclei. Four more events containing two $^3$H nuclei each were recorded. These can also be associated with $^6$Li charge exchange into the final state featuring a charge of two.\par

\indent We detected seven events in which the total charge of the fragments was equal to four. Of those, four were $^4$He+$2^1$H and one was $^3$He+$^2$H+$^1$H; these can be associated with the symmetric $^6$Li charge exchange resulting in formation of $^6$Be followed by its decay. In the remaining events, the process changing the projectile charge produces more particles in the final state. According to these results, the probabilities of the transitions in which a $^6$Li projectile goes over to states with charges $Z=2$ and 4 are about 1\% each, while the probabilities of meson-pro duct ion-free charge-exchange processes do not exceed 0.5\%.\par

\section{\label{sec:level8}COHERENT DISSOCIATION OF $^6$Li NUCLEI}

\indent The transverse momenta of nuclear fragments in the reference frame comoving with the nucleus undergoing fragmentation are determined by the momenta of Fermi motion of intranuclear nucleons. In an interaction with a target nucleus, however, the fragmenting projectile nucleus as a discrete unit can receive additional momentum. In this case, the fragment transverse momentum measured in the laboratory frame is not equal to the fragment momentum in the nucleus undergoing fragmentation. It is difficult to take into account the momentum received by the nucleus because the neutrons appearing to be fragmentation products are not recorded experimentally. If, however, the nucleus dissociates only into charged fragments, in which case the sum of the fragment mass numbers is equal to the mass number of the original nucleus, we can determine both the transverse-momentum transfer to the projectile nucleus and the fragment transverse momenta in the reference frame comoving with the nucleus undergoing fragmentation.\par

\indent Below, we consider separately the processes of $^6$Li coherent dissociation that are not accompanied by the production of relativistic charged particles. The topology of these events is illustrated in Table~7, which lists the numbers of $^6$Li dissociation events with and without the excitation of the projectile nucleus ($N_h\ne0$ and $N_h=0$, respectively).\par

\begin{table}
\caption{\label{Table:7} Number of events of $^6$Li coherent dissociation)}
\label{Table:7}       
\begin{tabular}{c|c|c}
\hline\noalign{\smallskip}
 & \multicolumn{2}{c}{Numder of events}\\
Dissociation & without the excitation & with the excitation\\
channel& of the target nucleus & of the target nucleus \\
 & ($N_h$=0) & ($N_h\ne0$)\\
\noalign{\smallskip}\hline\noalign{\smallskip}
$^4$He+$d$ & 23 & 24\\
$^3$He+$t$ & 4 & 1\\
$t+d+p$ & 4 & 3\\
$d+d+d$ & 0 & 2\\
\noalign{\smallskip}\hline
\end{tabular}
\end{table}

\begin{figure}
    \includegraphics[width=4in]{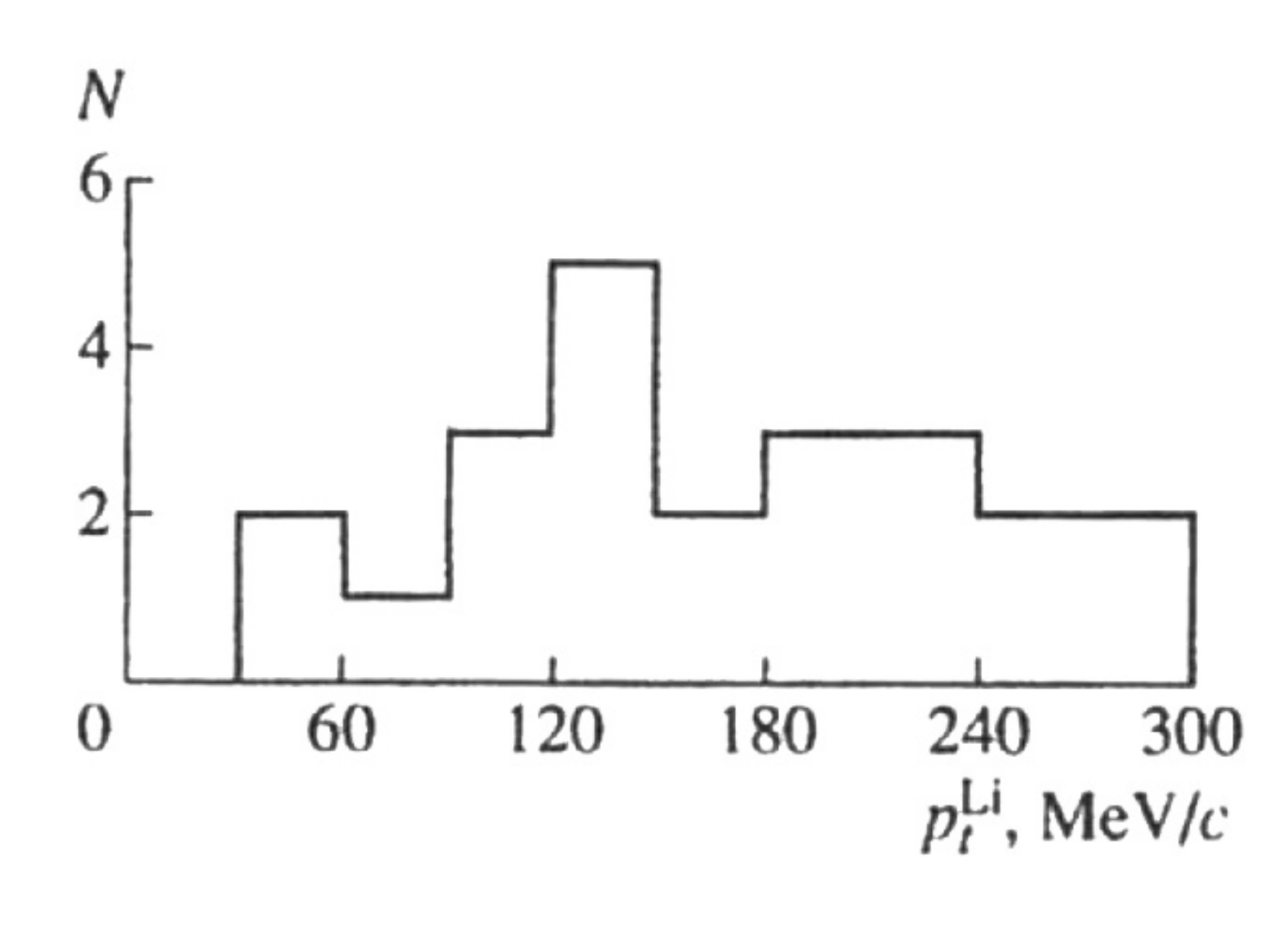}
    \caption{\label{Fig:5} Distribution of events with respect to the $^6$Li transverse momentum for the coherent-dissociation channel $^6$Li$\rightarrow\alpha+d$.}
    \end{figure}
	
\begin{figure}
    \includegraphics[width=4in]{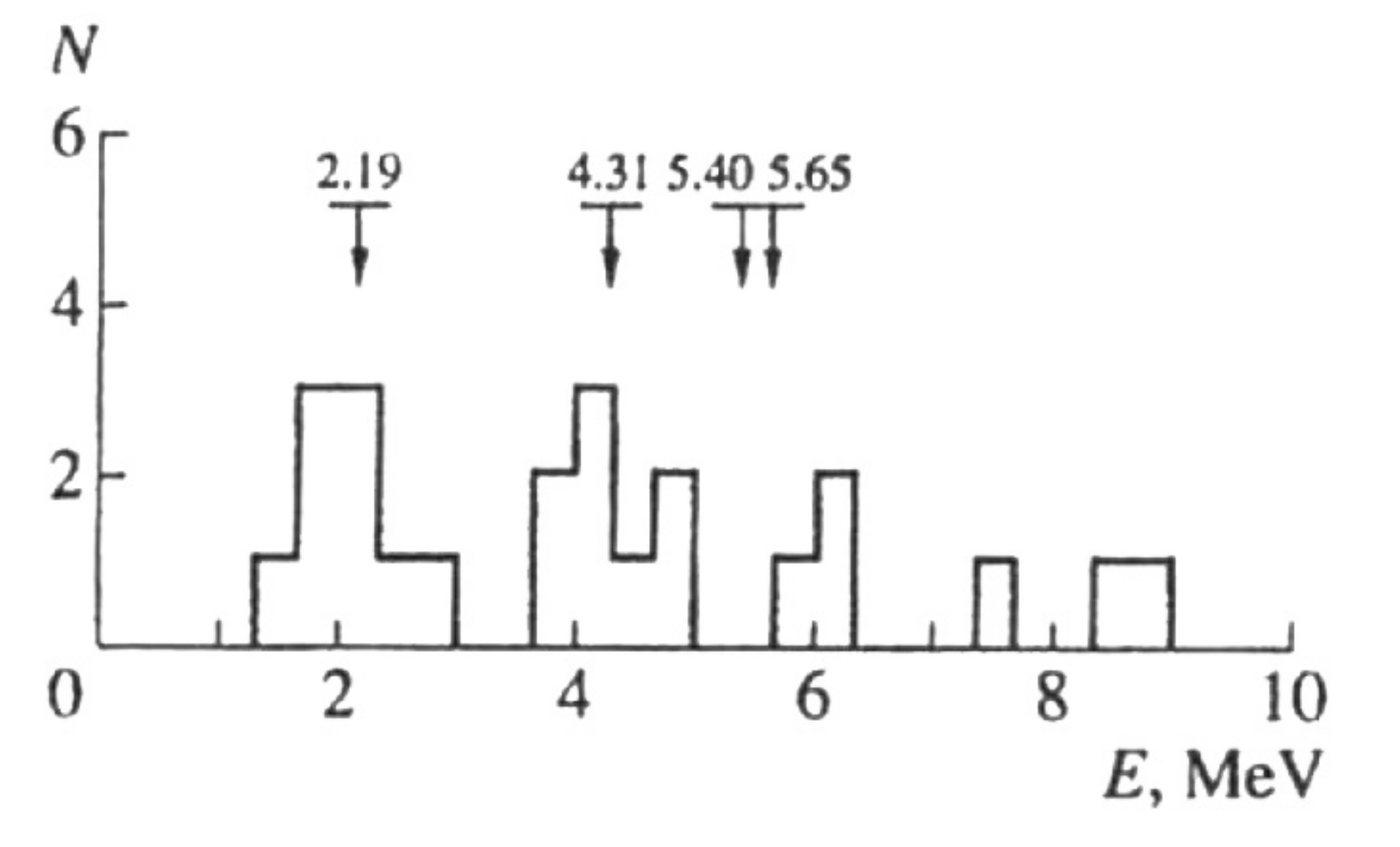}
    \caption{\label{Fig:6} Distribution of events with respect to the $^6$Li excitation energy in the coherent-dissociation channel $^6$Li$\rightarrow\alpha+d$. The arrows indicate the known levels of $^6$Li excitation.}
    \end{figure}

\indent Let us now investigate the transition $^6$Li$\rightarrow\alpha+d$ not accompanied by the excitation of the target nucleus; there are the vastest statistics for this channel. In the laboratory frame, the mean transverse momenta of the fragments in this channel are $\langle p^{\alpha}_t\rangle=150\pm14$~MeV/$c$ and $\langle p^d_t\rangle=134\pm16$~MeV/$C$. In each event, we determined the transverse-momentum transfer to the $^6$Li nucleus as the vector sum of the transverse momenta of individual fragments, $p^{Li}_t=p^{\alpha}_t+p^d_t$, and, by using the result obtained for $p^{Li}_t$ the transverse momenta of the fragments in their c.m. frame. Figure 5 displays the $p^{Li}_t$ distribution of events. The width of the interval of the transverse-momentum transfers to the $^6$Li nucleus does not exceed 300~MeV/$c$, while its mean value over all events is 164$\pm$15~MeV/$c$. The mean transverse momentum of the fragments in their c.m. frame is $\langle p^{\alpha}_t\rangle=\langle p^d_t\rangle=97\pm9$~MeV/$c$, which is considerably less than those measured in the laboratory frame. Such low a value of the transverse momenta is associated with the large radius of the two-cluster structure of the $^6$Li nucleus. In each event, the fragment transverse momenta were used to determine the effective mass $M_{eff}$ of the $\alpha+d$ system without taking into account the possible changes in the fragment longitudinal momenta. The difference of the effective mass $M_{eff}$ and the mass of the $^6$Li nucleus, $E=M_{eff}—M_{Li}$, can serve as an estimate of the $^6$Li excitation energy. Figure 6 shows the distribution of events with respect to the quantity $E$. The arrows indicate the known excitation levels of $^6$Li in this region (2.19, 4.31, 5.40, and 5.65 MeV). It can be seen that, in the distribution that we obtained, the majority of events populate the region of low-lying levels; that is, the dissociation process through the channel being investigated proceeds predominantly via the excitation of low-lying levels of $^6$Li. So unusually strong a reduction of the energy threshold for inelastic interactions must lead to a corresponding increase in the contribution to the total cross section from the interactions with large collision parameters and in the electromagnetic-interaction contribution. Such interactions are characterized by low momentum transfers. In accordance with this, the transverse-momentum transfer to $^6$Li nuclei does not exceed 200~MeV/$c$ in half of coherent-dissociation events; this is peculiar to electromagnetic interactions in nucleus-nucleus collisions.\par

\section{\label{sec:level9}CONCLUSION}

\indent The isotopic composition of fragments and the channels of $^6$Li fragmentation that are observed in the inelastic interactions of relativistic $^6$Li nuclei with photoemulsion nuclei are determined, to a considerable extent, by the loosely bound $(\alpha+d)$ structure of the $^6$Li nucleus. This cluster structure is responsible for the additional contribution to the cross section from inelastic interactions with large collision parameters and for an increased contribution of electromagnetic interactions. As a result, the mean range of $^6$Li nuclei with respect to inelastic interactions becomes smaller than that which would follow from the dependence on the projectile mass number. The mean-range value that we obtained corresponds to the effective radius that is close to the root-mean-square radius of the charge distribution in the $^6$Li nucleus. Our results also indicate that the pattern of $^6$Li interactions with nuclei is strongly affected by events in which individual loosely bound $^6$Li clusters interact with target nuclei.\par

\indent We have recorded only the channels of $^6$Li dissociation into charged fragments. It has been established that low-lying levels of $^6$Li excitation are manifested in the dissociation process.\par

\indent According to our results, the probabilities of $^6$Li charge exchange on photoemulsion nuclei into states with charges $Z=2$ and 4 via meson-production-free processes do not exceed 0.5\%.\par

\begin{acknowledgments}
\indent We are grateful to the staff of the Laboratory of High-Energy Physics at JINR (Dubna) for placing emulsion stacks at our disposal; to F.G. Lepekhin, B.B. Simonov, and M.M. Chernyavskii for discussion of the results; and to the members of a collaboration studying inelastic interactions of $^4$He and $^{12}$C nuclei at a momentum of 4.5~GeV/$c$ per projectile nucleon for their kind permission to use their experimental data in our analysis.\par
\indent This work was supported by the Russian Foundation for Basic Research.\par
\end{acknowledgments}

\end{document}